\documentclass[aip,reprint, graphicx]{revtex4-1}
\usepackage{graphicx}
\usepackage{textgreek}
\draft 

\usepackage[colorinlistoftodos]{todonotes}

	\usepackage{fancyhdr}
\begin{document}


\title{Local non-linear excitation of sub-100 nm bulk-type spin waves by edge-localized spin waves in magnetic films} 



\author{Pawel Gruszecki}
\email[]{gruszecki@amu.edu.pl}
\affiliation{Faculty of Physics, Adam Mickiewicz University, ul. Uniwersytetu Poznanskiego 2, 61-614 Poznan, Poland}

\author{Igor L. Lyubchanskii}
\affiliation{Donetsk Institute for Physics and Engineering named after O. O. Galkin
(branch in Kharkiv) of the National Academy of Sciences of Ukraine,
03028 Kyiv, Ukraine}
\affiliation{V. N. Karazin Kharkiv National University, 61022 Kharkiv, Ukraine}

\author{Konstantin Y. Guslienko}
\affiliation{Division de Fisica de Materiales, Depto. Polimeros y Materiales Avanzados: Fisica,
Quimica y Tecnologia, Universidad del Pais Vasco, UPV/EHU, Paseo M. Lardizabal 3,
20018 San Sebastian, Spain}
\affiliation{IKERBASQUE, the Basque Foundation for Science, Plaza Euskadi 5, 48009 Bilbao, Spain}

\author{Maciej Krawczyk}
\affiliation{Faculty of Physics, Adam Mickiewicz University, ul. Uniwersytetu Poznanskiego 2, 61-614 Poznan, Poland}


\date{\today}

\begin{abstract}
The excitation of high-frequency short-wavelength spin waves is a challenge limiting the application of these propagating magnetization disturbances in information processing systems. We propose a method of local excitation of the high-frequency spin waves using the non-linear nature of magnetization dynamics. We demonstrate with numeric simulations that an edge-localized spin wave can be used to excite plane waves propagating obliquely from the film's edge at a doubled frequency and over twice shorter in wavelength. The excitation mechanism is a direct result of the ellipticity of the magnetic moments precession that is related to the edge-mode propagation. As a consequence, the magnetization component tangential to the equilibrium orientation oscillates with doubled temporal and spatial frequencies, which leads to efficient excitation of the plane spin waves. Threshold-less non-linear process of short-wavelength spin-wave excitation proposed in our study is promising for integration with an inductive or point-like spin-torque source of edge spin waves. 

\end{abstract}


\maketitle 

Magnetic excitations in ferromagnetic materials propagate in the form of precessional coherent magnetization disturbances with microwave frequencies referred to as spin waves (SWs)\cite{gurevich1996book}. SWs are characterized by interesting properties, both from fundamental as well as applied point of view\cite{Barman2020,chumak2015magnon}. For example, in thin in-plane magnetized ferromagnetic films, wavelength of SWs can be several orders of magnitude shorter than those of electromagnetic waves of the same frequencies, which is crucial for miniaturization of microwave devices utilizing SWs. Furthermore, due to strong dipolar interactions the wavelength depends significantly on the SW  propagation direction, which allows, in particular, to form caustic SW  beams\cite{gieniusz2017switching}. The dipolar interactions create also nonuniformities of the internal magnetic field near the film's edges, which affect SW propagation\cite{davies2015towards,gruszecki2018mirage,gruszecki2014goos} and, moreover,  can form a potential well within which the localized SW modes can propagate\cite{sebastian2013nlESW,Davies2017}. The non-linear interaction of localized modes with incident non-localized SWs has recently drawn attention of the  magnonic community\cite{dadoenkova2019inelastic,zhang2018DMI_3magnon}.
Another consequence of the dipolar interactions is an ellipticity of the precession of magnetic moments due to shape anisotropy\cite{gurevich1996book}. In a thin layer this is manifested by a larger amplitude of oscillation of the transverse magnetization component, which is aligned in the plane of the film, than the out-of-plane component. Moreover, since the magnetization vector length is conserved, the magnetization component directed along the equilibrium direction can  contribute to oscillations, but with doubled frequency. Thus, the ellipticity of magnetic moment precession causes nonlinear phenomena to occur. This phenomenon is used in the parallel parametric pumping process\cite{schlomann1960,BRACHER2017}, where SWs at a particular frequency are excited by a homogeneous microwave magnetic field of twice higher frequency being directed parallel to the equilibrium magnetization orientation.

Oscillation of the longitudinal magnetization component at a doubled frequency, and related oscillation of the demagnetizing field, have been found responsible for the generation of the SWs at a doubled frequency of the original excitation\cite{Demidov2011,Rousseau2014}. The process observed with Brillouin light scattering measurements in thin ferromagnetic waveguides showed that symmetric SW excites non-resonantly antisymmetric wave in a threshold-less non-linear process\cite{Demidov2011}. In the case of the off-resonance excitation of the primary oscillations in the elliptical-shaped thin nanodot with homogeneous microwave field, the generation of the resonant \textit{2nd} and \textit{3rd} harmonics with antisymmetric and symmetric profile across the waveguide width, respectively, were observed by Demidov et al.~\cite{Demidov2011b}. Higher harmonics were also detected in resonantly excited microwave strip in broad-band ferromagnetic resonance measurements\cite{Marsh2012}. The excitation of the even and odd-order harmonics was related to the quadratic, dependent on longitudinal components of the magnetization, and cubic dependent on the transverse components, terms in the nonlinear Landau-Lifshitz equation, respectively.  However, the usefulness of the higher harmonic generation of SWs for magnonics, has not yet been demonstrated.

One of the vital challenges in miniaturization of magnonic circuits, expected for the next generation of the information and communications technology \cite{IRDS} limiting their practical application\cite{mahmoud2020introduction}, is the excitation of sub-100 nm wavelength SWs. A transducer directly converting microwave currents into short-wavelength SWs would be especially effective, but due to the limitations related to the downsizing of microwave antennas\cite{ciubotaru2016, gruszecki2016microwave},  alternative techniques are used.
These include utilizing grating couplers consisting of microwave antenna supplemented by a periodic array of magnetic elements placed underneath\cite{yu2013gratingCoupler, yu2019gratting, baumgaertl2020grCplrXRAY}, local resonances excited by a global  microwave magnetic field in the regions of magnetization or internal magnetic field inhomogeneities, as domain walls or film edges\cite{wiele2016,mushenok2017,whitehead2017, trager2019}. Quite recently also spin-transfer torque or spin-orbit torque driven excitations of SWs were also demonstrated \cite{Houshang2018,fulara2019}.

In this Letter, we demonstrate a method of resonant excitation of  ultra-short SWs in thin ferromagnetic film with the edge-localized SWs (E-SWs)\cite{lara2013eSWsTriangles,lara2017information} excited by the point source or microwave strip. We show with micromagnetic simulations, that at the edge of the film, the elliptical precession of magnetization of the propagating edge mode causes the spatial and temporal oscillation of the longitudinal magnetization component at doubled frequency with respect to the microwave field frequency. It enables the  excitation of the non-localized, "bulk-SWs" (B-SWs) propagating in the film's plane that become plane waves far away from the film edge. These waves propagate outward of the film's edge under the angles defined by the momentum conservation rule. The proposed process allows for the significant down conversion of the wavelengths of the edge excited SWs, broadband functionality and effective tuning.

\begin{figure}
\includegraphics[width=8.6cm]{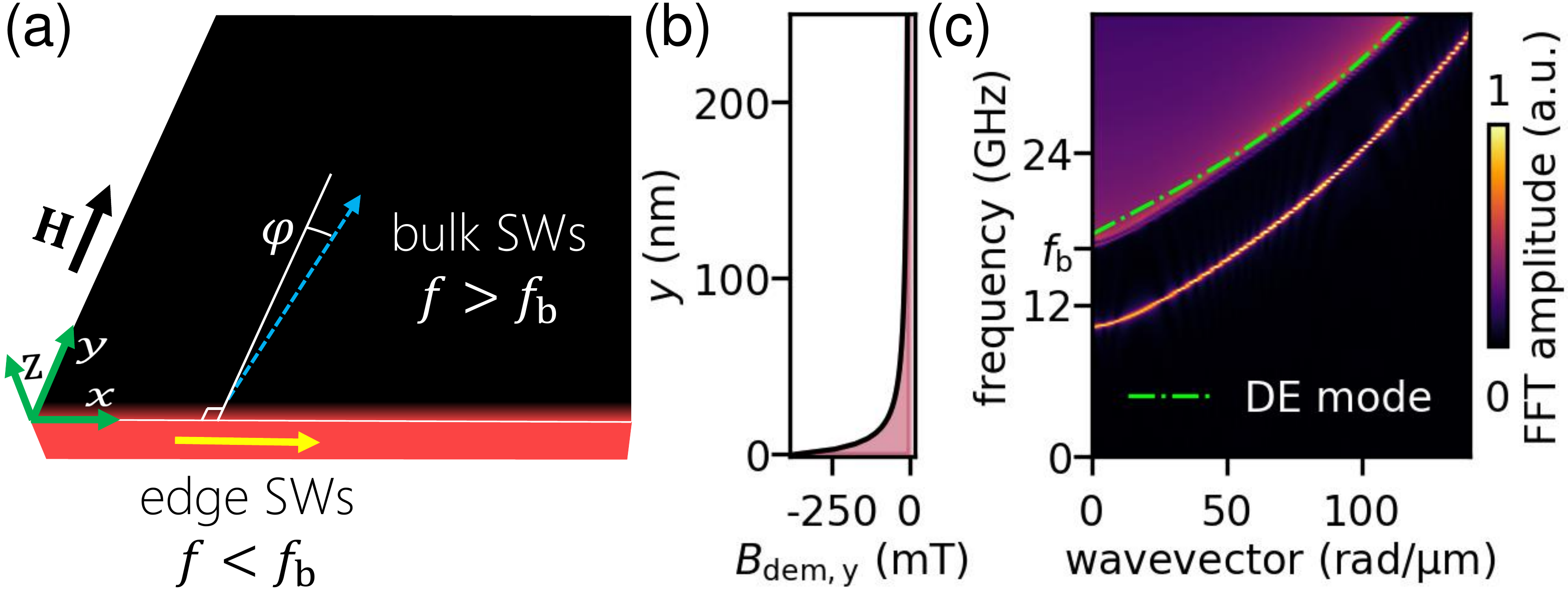}
\caption{\label{fig:Fig1} 
(a) Scheme of the system under consideration. 
(b) The profile of the demagnetizing field along the $y$-axis, normal to the film edge. 
(c) Simulated dispersion relation of SWs (colormap in background) and analytical dispersion of Damon-Eshbach mode in the film (the dash-dotted green line) calculated using Eq.~(\ref{eq:Dispersion}).}
\end{figure}

Let us consider a semi-infinite  permalloy (Py) film of thickness $d=10$ nm, that is magnetized in the film plane perpendicularly to its edge (Fig.~\ref{fig:Fig1}(a)). Due to dipolar interactions, a static demagnetizing field locally lowers the value of the effective magnetic field near the edge (see, Fig.~\ref{fig:Fig1}(b)). This local reduction of the internal field is utilized as a channel along which E-SWs, can propagate. This localization at the film edge distinguishes E-SWs from B-SWs spreading through the whole film. The dispersion relation for the investigated thin film with the saturation magnetization $M_\mathrm{S}=800$ kA/m, the exchange stiffness constant $A=13$ pJ/m, and the reduced damping constant $\alpha=10^{-4}$, being magnetized by the bias magnetic field of value $B_0=0.3$ T directed along the $y$-axis is calculated using mumax3 environment\cite{vansteenkiste2014design}. Details of simulations can be found in the supplementary materials. 
In these simulations, SWs are excited by the local point-source-like microwave pulse of the out-of-plane directed magnetic field located at the edge of film at the point $(x,y)=(0,0)$. The field  used for the excitation is described by the following function, $b_z (t;x,y)\propto p(x,y)\times \mathrm{sinc}[2\pi f_\mathrm{cut} (t-t_0 )]$, where $f_\mathrm{cut}=70$ GHz is the cut-off frequency, $t_0=10/f_\mathrm{cut}$ is the time-delay, and the point-source-like spatial profile of the microwave field is defined by the function $p(x,y)=\mathrm{exp}[-(x^2+y^2 )/\eta]$ with $\eta=40$ nm$^2$ describing its size. The resulting dispersion relation corresponds to the averaged over the $y$-coordinate the absolute value of two-dimensional fast Fourier transform (FFT) of the out-of-plane component of the reduced magnetization vector ($m_z$) over time and the $x$-coordinate for all  values of $y$ considered in the simulations, i.e., $D(f,k_x) =\left<|\mathrm{FFT}_{(t,x)} \{m_z(t;x,y)\}|\right>_y$  (we denote $m_i$ as the $i$-component of the reduced magnetization $m_i=M_i/M_\mathrm{S}$, where $i\in \{x,y,z\}$ and $m_i \in [-1, 1]$). In the obtained dispersion relation, shown in Fig.~\ref{fig:Fig1}(c), we can see the brightened parabolic area for $f>f_\mathrm{b}=16.5$ GHz ($f_\mathrm{b}$ is a bottom of B-SW band), which corresponds to the continuum of B-SWs. Besides, there is also a single frequency band, shifted downward from the B-SWs by circa 6 GHz, associated with the E-SWs. This 6 GHz wide gap separating both types of SWs at $k=0$ means that within the frequency range from 10.5 GHz  up to 16.5 GHz only the E-SWs can be excited.

\begin{figure}[!]
\includegraphics[width=8.6cm]{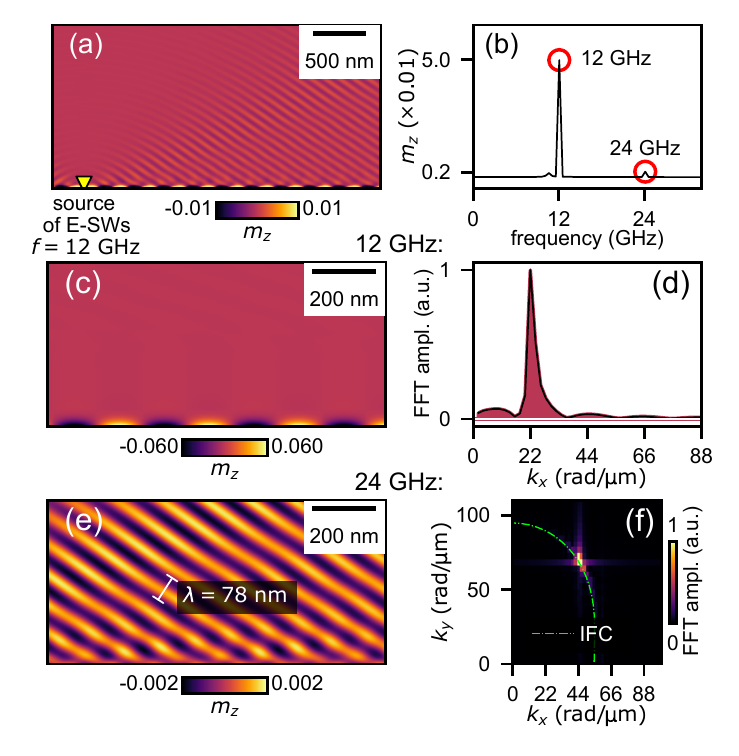}
\caption{\label{fig:Fig2} 
(a) Snapshot of the dynamical magnetization in the steady state with E-SW excitation at 12 GHz and (b) SW frequency spectrum obtained by micromagnetic simulations. In both figures the out-of-plane magnetization component ($m_z$) is used. The simulated SW mode profiles of $m_z$ at frequency 12 GHz (c) and 24 GHz (e) with the color precisely corresponding to the amplitude of $m_z$ at a given frequency. One- and two-dimensional fast Fourier transforms of SW modes at frequencies 12 GHz and 24 GHz are shown in (d) and (f), respectively. In (f) the bright spot corresponds to the dominating wavevector, the dash-dotted green line corresponds to the isofrequency contour (IFC) of the SW in thin film calculated at 24 GHz.  }
\end{figure}

To verify the possibility of the E-SW excitation, we excite magnetization dynamics by the out-of-plane microwave magnetic field in the form $b_z (t;x,y)=b_\mathrm{mf} p(x,y)\mathrm{sin}[2\pi ft]$, where $f=12$ GHz and $b_\mathrm{mf}=50$ mT. The corresponding result of micromagnetic simulations in the steady-state, i.e., after long, continuous excitation, is shown in Fig.~\ref{fig:Fig2}(a), where color corresponds to the amplitude of the $z$-component of the reduced magnetization, $m_z$. The position of the source of SWs, i.e., the point  $(x,y)=(0,0)$, is indicated by the yellow triangle. First of all, we can observe an intense, circa 20 nm wide, narrow beam of SWs propagating along the edge. This is an E-SW. We can also see  less intense plane waves propagating obliquely from the edge. Their  amplitude  is circa 20 times smaller than the maximal amplitude of E-SWs in the vicinity of the film edge. It is an unobvious  result since there are no available solutions for the B-SWs at the frequency of excitation. In order to explain the origin of these plane waves we determine the SW spectrum using FFT over time, see Fig.~\ref{fig:Fig2}(b). Namely, we present there, the maximal amplitude at a given frequency in the system,
$\mathrm{max}[m_z(f)]=\mathrm{max}\left[\mathrm{FFT}_t\{m_z(t;x,y) \}(x,y)\right](f)$.
Indeed, we can see two main peaks, the most prominent one corresponds to the frequency of the excitation (12 GHz), whereas, the second peak corresponds to its doubled frequency (24 GHz). There is also visible a small peak at a frequency around 10.5 GHz, which corresponds to the uniform resonant edge mode, that is excited during the initiation of the SW excitation. This spectrum confirms the generation of the second harmonic of the E-SW.

\begin{figure}
\includegraphics[width=8.6cm]{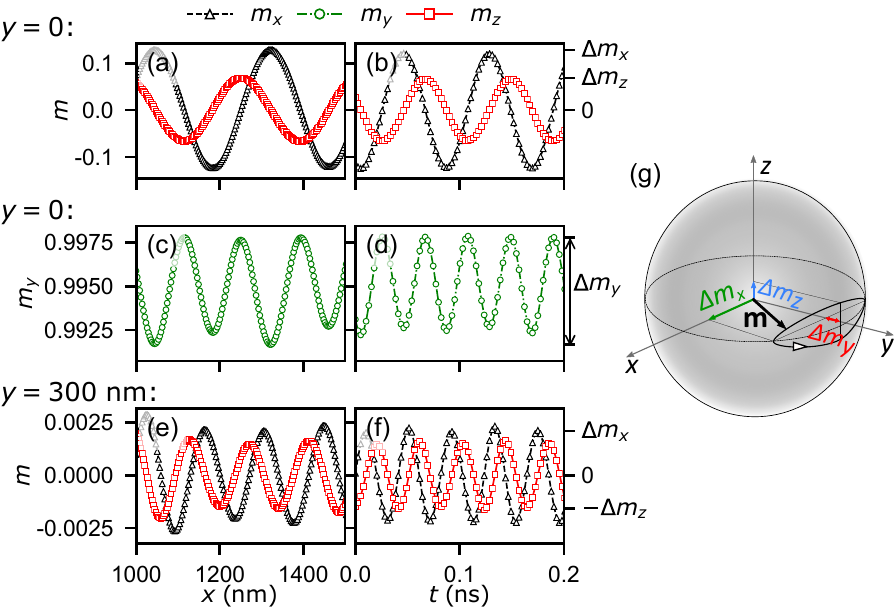}
\caption{\label{fig:Fig3} The dependencies of the transverse dynamic magnetization components $m_x$ and $m_z$ (red and dashed black lines) on $x$-coordinate and time at the film's edge ($y=0$) are depicted in (a), (b), whereas the longitudinal $m_y$-component dependence at the film's edge is shown in (c), (d). The oscillations of the $m_x$ and $m_z$ magnetization components (red and dashed black lines) at the distance $y=300$ nm from the film edge are shown in (e) and (f). (g) The cartoon that is schematically presenting the ellipticity of the magnetic moment precession.
}
\end{figure}

The mode profiles of the $m_z$ component of magnetization at 12 GHz and 24 GHz are presented in Fig.~\ref{fig:Fig3}(a) and (b), respectively. Indeed, the mode at 12 GHz corresponds to the E-SWs, whereas the mode at 24 GHz corresponds to the plane-wave B-SWs propagating obliquely outwards the film’s edge. The wavelength of the B-SWs is equal to $\lambda=78$ nm, whereas the wavelength of the E-SWs is equal to $\lambda_\mathrm{\text{e}}=285$ nm. It means that by  utilization of 285-nm-long E-SWs ($k_\mathrm{e}=22$ rad/\textmu m) we were able to excite 3.6 times shorter B-SWs at doubled frequency. This result is  similar  to the observation reported by Sebastian et al. \cite{sebastian2013nlESW}, where the second harmonic excitation of caustic beams, present only for the dipolar SWs, by the E-SWs in a 5 \textmu m-wide waveguide was observed. This is in contrast to our research, where we show the excitation of short, exchange-dominated SWs in a semi-infinite film propagating at the oblique direction.
Another similar result was reported by Hermsdoerfer et al. \cite{hermsdoerfer2009} who used the oscillations of the domain wall  to excite B-SWs with doubled the frequency of the domain wall oscillation.

The conversion of wavelengths can be better understood by transforming the real-space profiles of modes at frequencies 12 GHz and 24 GHz into the wavevector space, see Fig.~\ref{fig:Fig2}(d) and (f), respectively. We see that the SW at frequency 12 GHz in $k$-space is represented by a single peak at $(k_x,k_y)=(22$~rad/\textmu m$, 0)$ corresponding to the propagation of 285 nm long waves along the $x$-axis. In the case of 24 GHz mode we can also distinguish a single peak at (44 rad/\textmu m, 68 rad/\textmu m). It means that the value of tangential component of the B-SW wavevector  ($k_x$) is doubled with respect to E-SWs, $k_x=2k_\mathrm{e}$. The value of the $k_y$, on the other hand, must take the value to satisfy the SW dispersion relation of the thin film. The available wavevectors at a given frequency are represented by the isofrequency contour [IFC, marked by the green dash-dotted line in Fig.~\ref{fig:Fig3}(d)] being a cross-section of the dispersion relation surface at the selected frequency. In order to calculate IFC at $f=24$ GHz we use the analytical formula for dispersion relation of SWs in the in-plane magnetized thin film derived by Kalinikos and Slavin \cite{kalinikos1986theory}:

\begin{equation}
f=\frac{1}{2\pi}\sqrt{\left(\omega_{\mathrm{H}}+l_{\mathrm{ex}}^{2}\omega_{\mathrm{M}}k^{2}\right)\left(\omega_{\mathrm{H}}+l_{\mathrm{ex}}^{2}\omega_{\mathrm{M}}k^{2}+\omega_{\mathrm{M}}F\left(\varphi,k\right)\right)},\label{eq:Dispersion}
\end{equation}

where $\mu_{0}$ is the permeability of vacuum,  $l_{\mathrm{ex}}=\sqrt{2A/(\mu_{0}M_{\mathrm{S}}^{2})}$ is the exchange length, $\omega_{\mathrm{H}}=\left|\gamma\right|B_0$, and $\omega_{\mathrm{M}}=\gamma\mu_{0}M_{\mathrm{S}}$, and $\varphi$ is the angle  between the wavevector and the external magnetic field. The term $F(\varphi, k)$ is defined as:
\begin{eqnarray*}
F(\varphi,k)  =   1-P\left(k\right)\cos^{2}\varphi  
  +  \frac{M_{\text{S}}P\left(k\right)\left[1-P\left(k\right)\right]}{B_{0}\mu_0^{-1}+M_{\text{S}} l_{\mathrm{ex}}^{2}k^{2}}\sin^{2}\varphi,
\end{eqnarray*}
where  $P\left(k\right) = 1-\left[1-\mathrm{exp}(-k d)\right]/(k d)$.

Let us look at this phenomenon from a more classical point of view and compare the magnetization oscillations at the very film edge (at $y=0$) where E-SWs propagates, and along the line positioned 300 nm from the edge, where only B-SW can be present. For $y=0$, the $x$- and $t$-dependence of the transverse $m_x$ and $m_z$ magnetization components are presented in Fig.~\ref{fig:Fig3} (a) and (b), respectively, whereas   the $x$- and $t$-dependence of the longitudinal quasistatic $m_y$ magnetization component are presented in Fig.~\ref{fig:Fig3} (c) and (d). Due to the high amplitude of the E-SWs, the $y$-component of magnetization being aligned with the direction of the static external magnetic field oscillates with twice higher spatial and temporal frequency. This nonlinear effect is due to the ellipticity of the magnetization precession in thin film. Interestingly, when we plot the $m_x$ or $m_z$ transverse components of magnetization  at $y=300$ nm (see Fig.~\ref{fig:Fig3} (e) and (f)), we obtain oscillations along the $x$-axis with the same spatial and temporal frequencies, as respective $m_y$ oscillations presented in Fig.~\ref{fig:Fig3} (c) and (d)) (it is consistent with SW profiles presented in Fig.~\ref{fig:Fig2}). Importantly, the peak-to-peak magnitudes of these magnetization component oscillations are relatively similar, i.e., 0.006 vs. 0.004. It means that the oscillations of the quasi-static $m_y$ component of magnetization at the film edge are responsible for the excitation of the propagating bulk SWs, since they have the same frequency and wavevector component $k_x$. 
Moreover, comparing the magnetization precession components $m_x$ and $m_z$ for $y=0$ and $y=300$ nm, it can be observed that the ellipticity of magnetic moment precession is bigger for E-SWs than for B-SWs, i.e., $\mathrm{max}[m_z]/\mathrm{max}[m_x]=0.56$ vs. $0.72$, respectively.

\begin{figure}
\includegraphics[width=8.6cm]{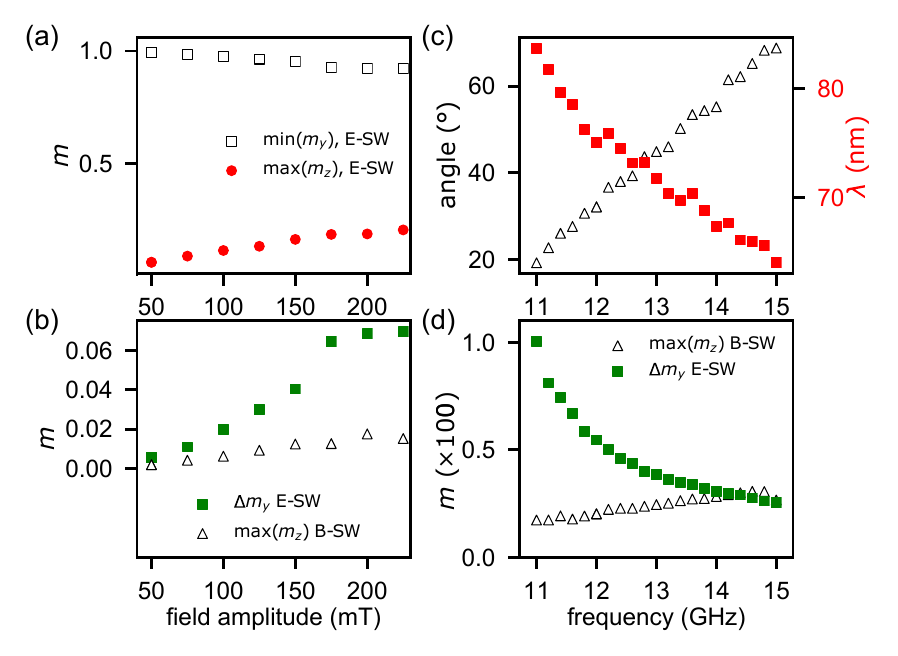}
\caption{\label{fig:Fig4} (a) Minimal amplitude of the dynamic $y$-component of the magnetization ($m_y$, empty black squares) and the maximal amplitude of the dynamic $z$-component of the magnetization ($m_z$, red dotes) at the film edge ($y=0$) as a function of the microwave field amplitude at frequency 12 GHz. (b) Maximal amplitude of the dynamic $z$-component of the B-SWs at $y=300$ nm (empty black triangles) and the amplitude of oscillations of the $y$-component of the magnetization at the edge
($\Delta m_y = \mathrm{max}[m_y(y=0)]-\mathrm{min}[m_y(y=0)]$),
green squares) plotted vs. the microwave field amplitude at frequency 12 GHz. (c) The propagation angle $\phi$ (empty black triangles) and wavelength (red squares) of the excited B-SWs vs. the frequency of the E-SWs excited by the microwave field of amplitude of 50 mT. (d) Maximal amplitude of the dynamic $z$-component of the B-SWs at $y=300$ nm (empty black triangles) and the amplitude of oscillations of the $y$-component of the magnetization at the edge
($\Delta m_y(y=0)$),
green squares) plotted vs. frequency of the E-SWs excited by the microwave field of amplitude of 50 mT.
}
\end{figure}

In order to verify how the emission of B-SWs depends on the amplitude and frequency of the  microwave field exciting E-SWs, we have performed a series of micromagnetic simulations with the microwave point-source. Firstly, we have studied how the amplitude of E-SWs and B-SWs depends on the amplitude of the microwave field at frequency 12 GHz, see Fig.~\ref{fig:Fig4}(a) and (b), respectively. Bigger amplitude of the microwave field results in bigger amplitude of the E-SWs and B-SWs, which is true up to 180 mT. 
In this range the the ratio the output B-SW intensity  to the input microwave field intensity reaches ca. 0.2\%\footnote{The effectiveness has been estimated as the ratio of output intensity of SWs to the input intensity of the peak pumping field $(m_z M_\mathrm{S} )^2/((b_\mathrm{mw}/\mu_0 )^2$ for data taken from Fig.~\ref{fig:Fig3} and Fig.~\ref{fig:Fig4} a-b}.
For the microwave field amplitude larger than 180 mT the process saturates and at $y=0$ the amplitude of $m_z$ reduced magnetization component oscillations is set at the level 0.2, whereas the peak-to-peak magnitude of $m_y$ oscillations [$\Delta m_y (y=0)$] reaches circa 0.06. At the same time at $y=300$ nm the amplitude of the $m_z$ magnetization component oscillations reaches the value of almost 0.002. This means that the efficiency of the excitation of the B-SWs by the E-SWs is limited and the amplitude of E-SW and B-SW saturates above the threshold microwave magnetic field amplitude due to nonlinear parametric excitation of high-$k$ SWs\cite{Bauer2015}. 

The results of simulations showing how the parameters of the excited B-SWs depend on the frequency of microwave field at amplitude 50 mT are shown in Fig.~\ref{fig:Fig4}(c)-(d). As the excitation frequency increases, the angle $\varphi$ of B-SWs increases, i.e., with higher frequency the propagation of the B-SWs is more parallel to the film edge.. 
We note that for the frequencies within the range 11 -- 15 GHz the tunability of the angles of B-SW propagation in the range of 50$^\circ$  is achieved. As easy to predict, the wavelength of the excited B-SW decreases  with the rise of the frequency (85 nm at 11 GHz and 65 nm at 15 GHz). The excitation frequency increase results in a slight increase in amplitude of the B-SWs and significant decrease of the peak-to-peak amplitude of the component  $m_y (y=0)$ . The decrease of $m_y(y=0)$ is due to the decreasing efficiency of the E-SW excitation by the microwave source for shorter E-SW.

Although the presented results were obtained with the use of point-source of the microwave field, the additional simulations made with the microwave magnetic field profiles that approximate microstrip antenna oriented along the $y$-axis ($p(x)=\mathrm{exp}[-x^2/\eta]$ and $\eta=40$ nm$^2$) at 50 mT and 170 mT amplitudes are qualitatively very similar to the point-source results.  Regarding the propagation distance of the considered SWs for the assumed value of $\alpha=10^{-4}$, we estimated that the E-SW propagate up to 76 \textmu m at 12 GHz at low amplitude of microwave field. The distance decreases with increasing the amplitude and at 50 mT of the microwave field the E-SW decay length is around 20 \textmu m. It  points at the feasibility of the experimental realization of the nonlinear excitation of short B-SWs by microstrip generated E-SWs, as well as spin-torque point contacts.

Summarizing, we have presented a method that allows to excite sub-100 nm-long SWs in the form of traveling plane waves. These waves are excited in nonlinear process by E-SWs, which possess strong ellipticity of magnetic moments precession. This causes  the spatial and temporal frequency of the longitudinal magnetization component, normal to the edge, to be doubled with respect to the edge mode. The plane waves are excited by the oscillations of the quasi-static magnetization component. Their length and propagation direction is a direct consequence of the requirement for conservation of the tangential  to the edge component of the wavevector. This effect can be understood as a local nonlinear excitation of plane-wave SWs propagating in a thin film with edge-localized SW  modes. The high frequency and short wavelength of exciting plane waves, combined with a simplicity of the method and feasibility for miniaturization of the system size to sub-\textmu m range, promise usefulness of the proposed  mechanism for magnonic applications.  

\subsection*{Suplementary material}
See the supplementary material for the details of micromagnetic simulations.

\begin{acknowledgments}
The research leading to these results has received funding from the National Science Centre of Poland, project no. 2019/35/D/ST3/03729. 
I.L.L. acknowledges support by COST action under project CA17123 MAGNETOFON.
K.Y.G. acknowledges support by IKERBASQUE (the Basque Foundation for Science) and by the Spanish Ministerio de Ciencia, Innovacion y Universidades grant PID2019-108075RB-C3-3.
The simulations were partially performed at the Poznan Supercomputing and Networking Center (Grant No. 398).
\end{acknowledgments}

\subsection*{DATA AVAILABILITY}
The data that support the findings of this study are available from the corresponding author
upon reasonable request.

\bibliography{main.bbl}

\end{document}